\documentclass[useAMS,usenatbib]{mn2e}
\usepackage{epsfig}
\usepackage{natbib}
\usepackage{graphicx}

\title{Optical Intraday Variability Studies of Ten Low Energy Peaked Blazars }
\author[Rani et al.]
{Bindu Rani$^{1}$\thanks{E-mail: bindu@aries.res.in},
Alok C.\ Gupta$^{1}$,
U.\ C.\ Joshi$^{2}$,
S.\ Ganesh$^{2}$,
Paul J.\ Wiita$^{3}$
\\
$^{1}$Aryabhatta Research Institute of Observational Sciences (ARIES),
Manora Peak, Nainital -- 263129, India\\
$^{2}$Physical Research Laboratory, Navrangpura, Ahmedabad-380 009, India \\
$^{3}$Department of Physics, The College of New Jersey, P.O.\ Box 7718, Ewing,
NJ 08628, USA }

\begin{document}

\date{Accepted ....... Received  ......; in original form ......}

\pagerange{\pageref{firstpage}--\pageref{lastpage}} \pubyear{2010}

\maketitle

\label{firstpage}

\begin{abstract}
We have carried out optical (R band) intraday variability (IDV) monitoring of a sample of ten bright low energy 
peaked blazars (LBLs). Forty photometric observations,  of an average of $\sim 4$ hours each, were made between 2008 September and 2009 June  
using two telescopes in India.  Measurements with good signal to noise ratios were  typically obtained within  1--3 minutes,
allowing the detection of weak, fast variations using N-star differential photometry.
We employed both structure function  and discrete correlation function 
analysis methods to estimate any dominant timescales of variability and found that in most of the cases any such
timescales were  longer than the duration of the observation. The calculated duty cycle of IDV in LBLs during our observing 
run is $\sim$52$\%$, which is low compared to many earlier studies; however, the
relatively short periods for which each source was observed can probably explain this difference.   
We briefly discuss possible emission mechanisms for the observed variability.

\end{abstract}


\maketitle

\section{Introduction}
Blazars are a subclass of radio-loud  AGNs characterized by strong and rapid flux variability across the 
entire EM spectrum and strong polarization from radio to optical wavelengths. 
Microvariability, or intraday variability (IDV) is commonly  observed across much of the electromagnetic (EM) spectrum 
of AGNs but is particularly common in  blazars. A change of flux of  $\sim$1--15\% 
within a few minutes to hours reflect extreme physical conditions embedded in small sub-parsec scales. 
According to the usually accepted orientation based unified 
model of radio-loud AGNs, blazar jets usually make an angle $\leq 10^{\circ}$ to our line-of-sight \citep{
urry1995}.  The Doppler boosting of the jet emission means that most of what we see from blazars arises in those
jets.  These 
jet dominated AGNs  provide a natural laboratory to study the mechanisms of energy extraction from the vicinity 
of central supermassive black holes, the physical properties of jets and perhaps also accretion disks. 

The radiation of blazars across the whole EM spectrum is predominantly non-thermal. At lower frequencies (through 
the UV or X-ray bands) the mechanism is almost certainly synchrotron emission while at higher frequencies 
it is very probably dominated by Inverse-Compton (IC) emission \citep{sikora2001, krawczynski2004}.
The spectral energy distributions (SEDs) of blazars have a double-peaked structure \citep{fossati1998,
ghisellini1997}. Based on the location of the peak of their SEDs, blazars are often sub-classified into the 
low energy peaked blazars (LBLs) and high energy peaked blazars (HBLs); 
the first component peaks in the near-infrared (NIR)/optical in case of LBLs and in the UV/X-rays in HBLs, while 
the second component usually peaks at GeV energies in LBLs and at TeV energies in HBLs.

The study of variability is one of the most powerful tools for revealing the nature of blazars and probing the various 
processes occurring in them.  Based on their different timescales, the variability of blazars can be broadly divided
into three classes, IDV, short-term variability (STV), and long-term variability (LTV).
Variations in the flux of source up to a few tenths of magnitude over a time scale of a day or less is known as IDV
\citep{wagner1995} or microvariability, or intra-night optical variability.  Variations of days to a few months are
often considered to be STV, while those from several months to many years are usually called LTV \citep[e.g.,][]{gupta2004};
both of these classes of variations for blazars typically exceed $\sim$ 1 magnitude and can exceed even 5 magnitudes.
Over the last two decades, the optical variability of blazars has been extensively 
studied on diverse timescales \citep[e.g.,][and references therein]{heidt1996, sillanpaa1996a, sillanpaa1996b, gupta2004, gupta2008a, gupta2008b, 
gupta2008c, gupta2009, ciprini2003, ciprini2007}. 

There are several theoretical models that might be able to explain the observed variability over wide time-scales 
for all bands, with the leading contenders all variants of models based upon 
shocks propagating down relativistic jets \citep{marscher1985, qian1991, hughes1991, marscher1992, wagner1995}. Some of 
the variability may arise from helical structures, precession or other geometrical effects occurring within the jets  
\citep[e.g.,][]{camenzind1992, gopal1992} and some of the radio variability is due to extrinsic 
propagative effects \citep{rickett2001}.  Hot spots or other disturbances in or above accretion disks surrounding the 
black holes at the centres of AGNs  \citep[e.g.,][]{mangalam1993, chakrabarti1993}  are likely to play a key role 
in the variability of non-blazar AGNs and might provide seed fluctuations that could be advected into a rotating blazar 
jet and then be Doppler amplified.

Despite the large amount of information we have about blazars that is very briefly summarized above, we still lack sufficient understanding of basic parameters 
of the emission regions, such as jet composition, a quantitative assessment of beaming parameters, or the processes leading to the 
origin of shocks in the jets.  These physical quantities are obviously important in understanding the physics of jets and their emission 
regions and additional IDV studies leading to statistically valid pictures of many blazars can help constrain them.

In this paper we present the results of an extensive IDV studies of a sample of ten LBLs including six BL 
Lacs and four FSRQs.  The work presented here  is focused on intraday variations in the R 
passband magnitudes of these sources, which were the brightest blazars visible from ARIES, Nainital, India and PRL, Mount Abu 
India. We also compare our observational results to some of those presented in the literature. 

The paper is structured as follows. In Section 2, we present the observations and data reduction procedure. Section 3 
provides our analysis and results. We present our discussion and conclusions in Section 4. 

\section{Observations and data reduction}
We have carried out optical R band photometric observations of ten LBLs from September 2008 to June 2009 
using two telescopes in India. The details of the telescopes and instruments used are given in Table 1 and the 
observation details are in Table 2.
 
For doing image processing, or data pre-processing, we generated a master bias frame for each observing night by
taking the median of all bias frames; the master bias frame was subtracted from all flat and
source image frames taken on that night. Then the master flat in each passband was generated by median
combine of flat frames in that passband. Finally, the normalized master flat in each passband was generated.
As usual, each source image frame was divided by the normalized master flat in the respective passband to remove 
pixel-to-pixel inhomogeneities (flat fielding). Finally cosmic ray removal was done from all source image
frames. This pre-processing of the data was done by using standard routines in the Image Reduction and Analysis
Facility\footnote{IRAF is distributed by the National Optical Astronomy Observatories, which are operated
by the Association of Universities for Research in Astronomy, Inc., under cooperative agreement with the
National Science Foundation.} (IRAF) software.

Our data analysis, or processing of the data, utilizes Dominion Astronomical Observatory Photometry (DAOPHOT II)
software to perform  aperture photometry \citep{stetson1987, stetson1992}. We carried out aperture 
photometry with four different aperture radii, 1$\times$FWHM, 2$\times$FWHM, 3$\times$FWHM and 4$\times$FWHM. On 
comparing the results, we observed that aperture radii = 3$\times$FWHM provided the best S/N ratio, and we have
adopted that in this work. For all of the 
ten blazars, we observed more than three local standard stars. The magnitudes of the standard stars we 
used in the fields of our sources are given in Table 3. The multiple comparison stars were used to check that the 
usual standard stars were non-variable. We have used two non-varying standard stars from each 
blazar field and plotted their differential instrumental magnitudes in Figs. 1$-$9. Finally, for the 
calibration of blazar data, we have used the one standard star that has a colour close to the blazar from those
two standards stars.  The calibrated light curves of the blazars are plotted in the same panel with the differential instrumental
magnitudes of two standard stars.

\section{Analysis and Results}
\subsection{Variability Parameters}
We checked for the presence of microvariability both by using the $F$-test \citep{deigo2010} and the variability 
detection parameter, $C,$ for the sake of comparison with earlier papers, which nearly always used that
approach. For two sample variances, say s$^{2}_{Q}$,
for the quasar differential light curves and s$^{2}_{stS}$, for that of the standard star
\begin{equation}
F =  \frac{s^{2}_{Q}}{s^{2}_{stS}}.
\end{equation}
The $F$-statistic is compared  with a critical value corresponding to the
significance level set for the test. We have used the inbuilt F test code available
in R\footnote{R: A language and environment for statistical computing. R Foundation
for Statistical Computing, Vienna, Austria. ISBN 3-900051-07-0, URL
http://www.R-project.org.}. A $p$-value of $\leq$ 0.01 ($\geq$99$\%$ significance level) is adopted for our 
variability detection criterion. 

The variability detection parameter is defined as  \citet[e.g.,][]{romero1999}) the average of $C1$and $C2$ where
\begin{eqnarray}
C1 = \frac{\sigma(BL - starA)}{\sigma(starA - starB)} \hspace{0.5 cm} \rm{and} \hspace{0.5 cm} C2 = \frac{\sigma(BL - starB)}{\sigma(starA - starB)}.
\end{eqnarray}
Here (BL $-$ starA) and (BL $-$ starB) are the differential instrumental magnitudes of the blazar and standard 
star A and the blazar and standard star B, respectively, while $\sigma$(BL$-$starA) and $\sigma$(BL$-$starB) 
are the observational scatters of the differential instrumental magnitude of the blazar and star A and the blazar 
and star B, respectively. If $C \geq 2.57$, a conservative confidence level of a variability detection is $> 99$\%, 
and we consider this to be a positive detection of a variation using this criterion. As noted by \citet{deigo2010} 
this C-statistic does not behave as a proper statistic should, but as it has been used in most of the IDV studies in 
literature,  we also employed this test. 

The percentage variation in the intraday light curves of LBLs is calculated by using the variability amplitude parameter, $A$, 
introduced by \citet{heidt1996}, defined as
\begin{eqnarray}
A = 100\times \sqrt{{(A_{max}-A_{min}})^2 - 2\sigma^2}(\%) ,
\end{eqnarray}
where $A_{max}$ and $A_{min}$ are the maximum and minimum magnitudes in the calibrated light curves of
the blazar and $\sigma$ is the average measurement error of the blazar light curve.
The calculated $F$-statistics, $C$-``statistics'' and variability amplitude parameter, $A$, values are listed in Table 4.

\subsection{Structure Function}
The first order structure function (SF) is a very useful tool to search for periodicities and timescales 
of variability in time series data trains \citep{simonetti1985}.
 Here we give only a very brief introduction to the method; for details refer
to \citet{rani2009}. The first order SF for a data train, $a$, is defined as
\begin{eqnarray}
D^{1}_{a}(k) = {\frac{1}{N^{1}_{a}(k)}} {\sum_{i=1}^N}  w(i)w(i+k){[a(i+k) - a(i)]}^{2},
\end{eqnarray}
where $k$ is the time lag, ${N^{1}_{a}(k)} = \sum w(i)w(i+k)$,
 and the weighting factor, $w(i)$
is 1 if a measurement exists for the $i^{th}$ interval, and 0 otherwise. 

The behaviour of the first order SF can be simply summarized.  The SF curves for AGN usually at first rise with time lag. 
Following  this rising portion, the SF will then fall into one of the following classes: (i) if no plateau exists, any 
time scale of variability exceeds the length of the data train; (ii) if there are one or more plateaus, each 
one indicates a possible time scale of variability; and (iii) if that plateau is followed by a dip in the SF, the lag
corresponding to the minimum of that dip indicates a possible periodic cycle (unless such a dip is seen at 
a lag close to the maximum length of the data train, when it is probably an artifact).  However, (iv) uncorrelated 
data produce a white noise behaviour, characterized by a constant slope \citep{ciprini2003}. 

We have carried out the SF analysis of all of those LCs which satisfy the variability detection criteria. 
Recently, some weaknesses of the SF method, including spurious indications of timescales and
periodicities have been discussed by \citet{uttley2010}, so we cross check the
SF results by using the discrete correlation function (DCF) method. The timescales of variability calculated using 
the SF analysis are listed in Table 4.

\subsection{Discrete Correlation Functions}  
The Discrete Correlation function (DCF) method was first introduced by \citet{edelson1988} and it was later 
generalized to provide better error estimates \citep{hufnagel1992}. Here we give only a brief introduction to the method;
for details refer to \citet{hovatta2007}, \citet{rani2009}, and references therein.

The first step is to calculate the unbinned correlation (UDCF) using the given time series through \citep{hovatta2007}
\begin{equation}
UDCF_{ij} = {\frac{(a(i) - \bar{a})(b(j) - \bar{b})}{\sqrt{\sigma_a^2 \sigma_b^2}}}.
\end{equation}
Here $a(i)$ and $b(j)$ are the individual points in two time series $a$ and $b$, respectively, $\bar{a}$ 
and $\bar{b}$ are respectively the means of the time series, and $\sigma_a^2$ and $\sigma_b^2$ are their variances.
The correlation function is binned after calculation of the UDCF. The DCF method does not automatically define 
a bin size, so several values need to be tried. If the bin size is too big, useful information is lost, but if 
the bin size is too small, a spurious correlation can be found. Taking $\tau$ as the centre of a time bin and $n$ 
as the number of points in each bin, the DCF is found from the UDCF via
\begin{equation}
DCF(\tau) = {\frac{1}{n}} \sum ~UDCF_{ij}(\tau) .
\end{equation}
The error for each bin can be calculated using
\begin{equation}
\sigma_{\mathrm def}(\tau) = {\frac{1}{n-1}} \Bigl\{ \sum ~\bigl[ UDCF_{ij} - DCF(\tau) \bigr]^2 \Bigr\}^{0.5} .
\end{equation}

A DCF analysis is frequently used for finding the correlation and possible lags between multi-frequency 
AGN data where different data trains are used in the calculation \citep[e.g.,][and references therein]{villata2004, raiteri2003, 
hovatta2007}. When the same data train is used, so $a$=$b$,
there is obviously a peak at zero DCF, indicating that there is no time lag between the two,
but any other strong peaks in the DCF give indications of variability timescales. The calculated $t_{v}$
values using DCF analyses are listed in Table 4.

\subsection{Intraday Variability (IDV) of Individual Blazars}
The R filter light curves (LCs) of the blazars in which significant variability has been detected, along with their 
corresponding SF and DCF analysis curves, are displayed in Figures 1$-$7; the remaining non-variable LCs 
of the blazars are plotted along with the corresponding curves of the  standard stars used for comparison
in Figures 8 and 9. The complete observing log for the blazars in given in Table 3. The values of $A$, $C$, $F$-test 
along with any estimated timescales of variability found using SF and DCF analysis methods on the individual blazar 
LCs are listed in Table 4. A detailed multiband optical short term variability (STV) study of the fluxes 
and colours of all of these blazars over the same time period of observation is reported in \citet{rani2010b}. 
There we showed that the colour versus brightness correlations seen in these sources support the hypothesis 
that BL Lacs tends to be bluer with increase in brightness while FSRQs shows the opposite trend. We now 
report some key individual results for each of our sources, placed in the context of earlier work. \\

\noindent
{\bf 3C 66A:} This is a low energy peaked blazar (LBL) at redshift $z = 0.444$ \citep{lanzetta1993} and 
belongs to the class of BL Lac objects. 
Since its optical identification \citep{wills1974}, the source has been regularly monitored at many 
observable frequencies, although less regularly at radio frequencies \citep{aller1992, takalo1996}. \citet{fan1999, fan2000} 
have studied the long-term optical and IR variability of the source and reported a variation of 
$\leq$1.5 mag at time scales of $\sim$1 week to several years at those two frequencies. \citet{bottcher2005} 
reported a large microvariability of $\sim$0.2 mag within 6 hours; they also reported 
several major outbursts in the source separated by $\sim$50 days and argued that the outbursts seem 
to have a quasi-periodic behaviour. At the end of 2007 the source was found to be in an optically active 
phase, which triggered a new Optical-IR-Radio Whole Earth Blazar Telescope (WEBT) campaign on the source 
\citep{bottcher2009}. 

Our optical IDV observations of the source 3C 66A comprise a total of seven LCs, spanning  a time period between
October 2008 and January 2009. During this period a change of $\sim$1 magnitude in brightness of the source is seen 
\citep{rani2010b}. The source showed significant microvariability only on October 22 and 26, 2008 (Fig.\ 1). There is 
a continuous fading trend of $\sim$0.08 and $\sim$0.06 magnitude, respectively on those two nights. The SF and 
DCF analysis of the LCs revealed that any timescale of variability in this source at those epochs is greater than 
the lengths of our observations. 

\noindent
{\bf AO 0235+164:} The blazar AO 0235+164 at $z = 0.94$ \citep{nilsson1996} was classified as a BL Lac object
by \citet{spinrad1975}. Over the past few decades this blazar has been seen to be highly variable over 
all timescales and at all frequencies  \citep{ghosh1995, heidt1996, raiteri2001} 
and a very high fractional polarization of $\sim$40$\%$ has been reported in the source both 
at visible and IR frequencies \citep{impey1982}. By analysing 25 years (1975$-$2000) of optical and radio 
data, \citet{raiteri2001} argued that the source seemed to have an optical outburst period of $\sim$5.70 years 
but the expected outburst in 2004 was not detected by a 2003--2005 multiwavelength WEBT observing campaign \citep{raiteri2006a}.  
A more detailed long term optical data analysis suggested a possible outburst period of $\sim$8 years 
in this source \citep{raiteri2006b} and this period was supported by subsequent observations \citet{gupta2008c}.  Strong 
IDV flux variations of 9.5$\%$ and 13.7$\%$ during two nights were observed by \citet{gupta2008c}.
Recently, \citet{rani2009} reported the possible presence of nearly periodic fluctuations, with a timescale of $\sim$17 days, in a
12.7 year long X-ray light curve of AO 0235$+$164 obtained by the All Sky Monitor (ASM) instrument on the Rossi X-ray Timing 
Explorer (RXTE) satellite.  

We observed three IDV LCs of the source AO 0235+164  between October 2008 and January 2009. The brightness 
of the source decayed by  $\sim$2.2 magnitude during this period \citep{rani2010b}. We found a 
significant flux variation in two out of three LCs. A continuous fading trend of $\sim$0.13 magnitude and 
both brightening and decaying of $\sim$0.04 magnitude were observed on October 20 (Fig.\ 1) and October 23, 2008 (Fig.\ 2), 
respectively. A possible timescale of variability is $\sim$5.2 hr (from the SF analysis) for the LC observed 
on 20 October, while any such timescale exceeds the length of our observation on 
23 October.      However this putative timescale is not supported by the DCF analysis.

\noindent
{\bf PKS 0420$-$014:} The blazar PKS 0420$-$014 is classified as a FSRQ and has a redshift of 0.915. It has 
been observed in optical bands since 1969.  Several papers have reported multiple optically active and 
bright phases of the source and perhaps regular major flaring cycles \citep[e.g.,][and references therein]{villata1997, webb1998, 
raiteri1998}. \citet{webb1998} reported that there were increases 
of $\sim$ 2$-$3 magnitudes during the active phases of this blazar during their  observations that stretched
from December 1969 to January 1986. \citet{clements1995} have reported variations of $\Delta$mag $\cong$ 2.8 mag 
with a time scale of $\sim$22 years. 

We found a variation of $>$0.1 magnitude within 2 hrs in the brief optical LCs of the source on both of the days 
of observation in October and December 2008. During this period the source  brightened by a factor of 
$\sim$0.7 magnitude \citep{rani2010b}. The nominal timescale of variability (from the peak of the SF) is 0.12 hrs for the LC 
observed on December 26 and multiple dips might indicate a possible periodicity around 0.18 hrs, 
which is weakly supported by the DCF curve.  The other LC is irregular and shows no hint of a timescale  (Fig.\ 2).  

\noindent
{\bf S5 0716$+$714:} The blazar S5 0716$+$714 is classified as a BL Lac object. \citet{nilsson2008} 
made a recent claim of redshift determination of the source to be $z = 0.31\pm0.08$.
This source has been 
extensively studied at all observable wavelengths from radio to $\gamma$-rays on diverse time scales 
\citep{wagner1990, heidt1996, villata2000, raiteri2003, montagni2006, foschini2006, ostorero2006, 
gupta2008a, gupta2008c}. This source is one of the brightest 
BL Lacs in optical bands with an IDV duty cycle of nearly 1.  Unsurprisingly, it has been the subject of 
several optical monitoring campaigns on IDV timescales \citep{heidt1996, montagni2006, gupta2008c}. This 
source has shown five major optical outbursts \citep{gupta2008c} at intervals of $\sim$3.0$\pm$0.3 years. 
High optical polarizations of $\sim$ 20$\%$ - 29$\%$ has also been observed in the source \citep{takalo1994, 
fan1997}. \citet{gupta2009} reported good evidence for nearly periodic oscillations in a
few of the intraday optical light curves of the source observed by \citet{montagni2006}. Good evidence of 
presence of a $\sim$15 minute periodic oscillation at optical frequencies has been reported by \citet{rani2010a}.  

Our optical IDV observations of  S5 0716$+$714 spans a time period from October 2008 to January 2009. 
The source brightened by a factor of $\sim$2 magnitude during this period \citep{rani2010b}. We found significant 
microvariability of $\sim$0.1 magnitude in three out of four LCs of the source. The LCs observed on December 24, 2008 
and January 03, 2009 show continuous fading trends trend of the order of $\sim$0.1 magnitude, though the
former is abrupt while the latter is gradual.  Both fading and brightening 
and fading trends of $\sim$0.05 magnitude were observed over just a few minutes on December 23, 2008. The calculated 
possible variability timescales are listed in Table 4, but the lack of agreement between the SF and DCF possibilities
leads us to discount their reality.

\noindent
{\bf PKS 0735$+$178:} The blazar PKS 0735$+$718 has been classified as a BL Lac object \citep{carswell1974}.
Papers concerning  its redshift determination  \citep{carswell1974, falomo2000} 
 had set a lower limit of   $z > 0.4$ and  
$z = 0.424$ was reported for the source using a HST snapshot image \citep{sbarufatti2005}.
Since its optical identification, the source has been extensively observed across the whole electromagnetic spectrum 
\citep{teransa2004, gu2006, gupta2008c, ciprini2007}. 
A periodicity of $\sim$14 years has been suggested to be present in the source 
using a century long, but still sparse, optical light curve \citep{fan1997}.  Optical variability on
IDV and STV timescales has  been observed for 0735$+$178 \citep{xie1992, massaro1995, 
fan1997, zhang2004, ciprini2007, gupta2008c}.  A significant amount of polarization  
($\sim$ 1$\%$ to 30$\%$) has been observed in the source both at optical and IR bands \citep{mead1990, takalo1991, 
takalo1992b, valtaoja1991a, valtaoja1993, tommasi2001}.   

Our IDV observations of the source PKS 0735$+$718 comprise four LCs taken between December 2008 and January 
2009. The brightness of the source changes by $\sim$0.6 magnitude during this 
period \citep{rani2010b}. We found significant microvariations of an order of $\sim$0.1 magnitude in three out 
of four observed LCs of the source (Fig. 4). The only conceivable hints of  timescales for 
 variability from the SFs are $\sim$2.3 hrs for the LC observed 
on December 28, 2008 and $\sim$0.58 hrs for January 04, 2009.  However, since both of these peaks are close
to the total lengths of the observations they are not likely to be real, nor  supported by DCF results.

\noindent
{\bf OJ 287:} The blazar OJ 287 at  $z = 0.306$ is one of the most extensively observed and best studied 
BL Lac objects. It is also among the very few AGN's for which more than 
a century of optical observations are available \citep{sillanpaa1996a, sillanpaa1996b, 
fan1998, abraham2000, gupta2008c, fan2009a}. Using the binary black hole model \citep{sillanpaa1988} for 
the long-term optical light curve of the source, an outburst with a predicted  $\sim$12 year period was 
detected in the source by the OJ-94 programme \citep{sillanpaa1996a, valtonen2008}. 
A very high optical polarization and variability in the degree and angle of polarization  has been also 
reported for OJ 287 \citep{efimov2002}. The observational properties of the source from radio to X-ray 
energy bands have been reviewed by \citet{takalo1994}. Recently, \citet{fan2009a}, reported large  
variations in the source of $\Delta$V = 1.96 mag, $\Delta$R = 2.36 mag, and $\Delta$I = 1.95 mag during 
their observations spanning 2002 to 2007.        

Our observations of OJ 287 span a 
period from December 2008 to January 2009, and during this time the brightness of the source changed significantly 
by  $\sim$1 magnitude \citep{rani2010b}. 
The source showed significant microvariations only in one out of three observed nightly LCs during which the brightness of the 
source continuously faded by $\sim$0.08 magnitude within 2 hrs (Fig.\ 5). The SF and DCF analysis showed that any
timescale of variability is longer than the timescale of observation.

\noindent
{\bf 3C 273:} The FSRQ 3C 273 was the first quasar discovered and has a redshift of 0.158 \citep{schmidt1963}.
It is categorized as a LBL \citep{nieppola2006} and its spectral energy distribution, correlations among different 
energy band flares and the approaching jet orientation have been extensively  studied at all wavelengths. There 
are many papers covering the observational properties of the source in the optical band 
\citep{angione1971, sitko1982, corso1985, corso1986, moles1986, hamuy1987, sillanpaa1991, valtaoja1991b, 
takalo1992a, takalo1992b, elvis1994, lichti1995, ghosh2000}.  An analysis of the optical light curve of 
3C 273 spanning over 100 years can be interpreted to suggest a LTV timescale of $\sim$13.5 years \citep{fan2001}. Recently,
the short-term optical variability and colour index properties of the source have been studied by \citet{dai2009}. 

Our IDV observations of the source 3C 273 span the period from December 2008 to April 2009 during which a total of 
eight LCs were obtained. There is no change in overall flux of the source during this period \citep{rani2010b}. This source  
showed microvariations (of  $\sim$0.05 magnitude) in only two out of eight observed LCs (Fig.\ 5).  On the night of
19 April 2009 there is a hint of a timescale of variability of $\sim$6 hrs from both the SF and DCF approaches.

\noindent
{\bf PKS 1510$-$089:} The blazar PKS 1510$-$089 is classified as a FSRQ and has $z = 0.361$. It also belongs 
to the category of highly polarized quasars. Significant optical flux variations in the source were first 
reported by \citet{lu1972} over a time span of $\sim$5 years. The historical light curve of 
the source shows a large variation of $\Delta$B = 5.4 mag during an outburst in 1948 after which it
faded by $\sim$2.2 mag within 9 days \citep{liller1975}.  Strong variations on IDV time scales also have
been reported for PKS 1510$-$089; e.g., $\Delta$R = 0.65 mag within 13 min. \citep{xie2001}, $\Delta$R = 2.0 mag 
in 42 min. \citep{dai2001}, $\Delta$V = 1.68 mag in 60 min. \citep{xie2002a}. In the optical light 
curves of this source, a few deep minima have been observed  on different days  \citep{xie2001, dai2001,
xie2002b}, that nominally correspond to a time scale of $\sim$42 min, though no more than 3 such
dips were ever seen in a single night.  Nonetheless, an eclipsing binary black hole model was actually proposed to 
explain the occurrences of these minima \citep{wu2005}. Other observations by this group have yielded a claim 
of another possible  time scale between minima of $\sim$89 min \citep{xie2004}. 

We carried out IDV observations of the source from April to June 2009. There is a large change in the brightness of the 
source during this period , with  $\Delta$R$_{mag}$ = 1.5 \citep{rani2010b}. We found significant microvariations of $\sim$0.05--0.08 
magnitude in three out of four LCs (Fig.\ 6). We found that any timescale of variability is larger than the timescale 
of observations, except perhaps for the LC observed on April 19, 2009 for which it is formally $\sim$0.6 hrs from the
SF curve and $\sim$00.5 hr from the DCF; however,
this LC has too few points to allow the production of a crisp SF or DCF, so this evidence is very weak.

\noindent
{\bf BL Lac:} The object BL Lac at $z = 0.069$ \citep{miller1978} is the archetype of its class. Observations
over the  past few decades have showed that its optical and radio emissions are highly variable
and polarized and the polarization at those widely separated  frequencies is found to be strongly correlated
 \citep{sitko1985}. It is among the very few sources for which more than 100 years of optical
data is available in the literature \citep{shen1970, webb1988, fan1998}. An optical
variation of $\Delta$B = 5.3 mag and a possible periodicity of $\sim$14 years has been reported 
for BL Lac by \citet{fan1998}. Very recently, \citet{nieppola2009} have studied the long term variability 
of the source at radio frequencies and generalized the shock model that can explain it.       

Our IDV observations of the source BL Lac were made between September 2008 and June 2009. The brightness 
of the source faded by $\sim$1.6 magnitude during this period \citep{rani2010b}.  BL Lac
faded by $\sim$0.05 magnitude within 1 hr of observation on September 04, 2008 which had a nominal variability 
timescale of $\sim$2.5 hrs according to both the SF and DCF plots (Fig.\ 7). 
A continuous  rising trend of $\sim$0.1 magnitude in the LC of the source was 
observed on June 21, 2009, and the calculated timescale of variability for the LC that night exceeds the time 
period of the observations.

\noindent
{\bf 3C 454.3:} A FSRQ at a redshift of 0.859, 3C 454.3  is among the most intense and 
variable sources. The source has been detected in the flaring state in July 2007 and July 2008 at 
$\gamma$-ray frequencies and those flares have been found to be well correlated with optical and 
longer wavelength flares \citep{ghisellini2007, raiteri2008, villata2007}. The 
long term observational properties of the source at optical and radio frequencies have been well 
studied through multiwavelength campaigns   \citep[e.g.,][]{villata2006, villata2007}. The IDV
of the source was recently studied by \citet{gupta2008c}. They have reported 
that the amplitude varied by $\sim$5 - 17$\%$ during their observing span. An extraordinary flaring 
activity above 100 MeV has been reported in the source in December 2009 \citep{striani2010}.    

We observed  two IDV LCs of the source on October 24 and 28, 2008. During the period the brightness 
of the source increases by $\sim$0.4 magnitude \citep{rani2010b}. The source showed significant 
microvariations of $\sim$0.13 magnitude on October 24, 2008 (Fig. 7) while no significant variations 
has been detected for the LC observed on October 28, 2008. The SF and DCF analysis revealed that any
timescale of variability exceeds the length of the observation on the first of those nights.

\section{Discussion and Conclusions}
We have carried out optical R band IDV observations of ten LBLs spanning over a time period of 2008 
September to 2009 June. The sources PKS 0420$-$014 and PKS 1510$-$089 were in faint states; 
AO 0235$+$164, BL Lac and 3C 454.3 were possibly in post-outburst states; S5 0716$+$714 and 3C 66A were 
in pre-outburst states, while PKS 0735$+$178 and OJ 287 were in some intermediate states during the 
observing run \citep{rani2010b}. In our search of microvariations in ten LBLs we found significant IDV 
in 21 out of 40 observed LCs; so the calculated duty cycle of IDV in LBLs during our observing run is 
$\sim$52$\%$. We performed the SF and DCF analysis to calculate the nominal timescales of variability; however
we found that in most of the cases this timescale of variability is longer than the length of observations and
in a substantial majority of the cases where the SF indicated a possible timescale the 
DCF did not support it. 

The blazar emission mechanism in the outburst state is quantitatively understood by relativistic shocks
propagating through a relativistic jet of plasma. In general, blazar emission in the outburst state is non-thermal 
Doppler-boosted emission from jets enhanced by that arising from shocks in the flows.
\citep{blandford1978, marscher1985, marscher1992}. The other models of AGNs
that can explain the IDV in any type of AGN are optical flares, disturbances or hot spots on the accretion disk 
surrounding the black hole of the AGN \citep[e.g.,][ and references therein]{mangalam1993}. Models based on the 
instabilities on the accretion disk could convinancibly yield  blazar IDV only when the blazar is in the very 
low state. When a blazar is in the low state, any contribution from the jets, if at all present, is very weak. 
So, we consider that the observed IDV in the sources AO 0235$+$164, BL Lac, 3C 454.3, S5 0716$+$714, 3C 66A, 
0735$+$178 and OJ 287 is almost certainly related to a shock propagating through a relativistic jet \citep{blandford1979, 
marscher1985}. Turbulence behind a shock propagating down such a jet is a very feasible way to explain the observed 
IDV \citep{marscher1996}. Since the blazars PKS 0420$-$014 and PKS 1510$-$089 were observed in relatively faint 
states, there is a chance that the observed optical IDV in these source may be because of hot spots or any other enhanced 
emission on the accretion disk \citep{mangalam1993}. 

In one source, 3C 273, we are unable to classify the state of the source since this blazar has been in an essentially steady 
state for last six years \citep{dai2009, rani2010b}. This source showed significant microvariations in two out of eight 
observed LCs and on both of days of observation the brightness of the source follows both rising and fading 
trends of $\sim$0.05 magnitude. Whatever may be the mechanism responsible for the origin of 
of microvariability in this source, it does  not seem to be strong enough to introduce day-to-day variations in the flux of the 
source. 

It is worth noting that an extrinsic mechanism can also be responsible for some of the observed IDV in blazars. 
Extrinsic IDV could be caused by refractive interstellar scintillation, which  is only relevant in low-frequency 
radio observations, or microlensing, which is achromatic.  We note that the blazar AO 0235$+$164 has two 
foreground galaxies at $z = 0.524$ and $z = 0.851$ \citep{cohen1987, nilsson1996, webb2000}. The flux of 
this source is contaminated and absorbed by foreground galaxies, the stars of which can act as gravitational 
microlenses. Thus the observed  optical IDV in AO 0235+164 could arise, at least partially, from  gravitational 
microlensing.  Using two separated telescopes to simultaneously observe 0716$+$714
\citet{pollock2007} showed that instrumental and atmospheric effects cannot account for the microvariations they
measured for that blazar.

We found significant IDV in 21 out of 40 observed LCs of ten LBLs; so the calculated duty cycle of IDV in LBLs 
during our observing run is $\sim$52$\%$. The average duration of our observation was 3.7 hrs per LC. The SF and DCF 
curves revealed that in $\sim$60$\%$ cases any timescale of variability is longer than the timescale of observations.
In quite a few cases there were hints of timescales in the data from the SF plots, but only a few of those hints were
supported by the DCF plots.

Extensive IDV studies of different subclasses of AGNs revealed that the occurrence of IDV in blazars observed 
on a timescale of $<$6 hrs is $\sim$60$-$65$\%$ and if the blazar is observed more than 6 hrs then the possibility 
of IDV detection is 80$-$85$\%$ \citep[][and references therein]{carini1990, gupta2005}. If we consider  observations
over days to months, i.e., at 
STV timescales then the observed duty cycle of variations is $>$92$\%$ \citep[e.g.][]{rani2010b} and at LTV timescales 
it is almost 100$\%$, which confirms that the probability of detection of variability in blazars rises with the  
duration of the observations. Although the duty cycle from our current observations is  less than that 
reported by \citet{gupta2005}, since the average duration of our observations is $<$4 hrs, 
this is unsurprising.

\section*{Acknowledgments} 
Work at PRL is supported by the Department
of Space, Govt. of India.

%\bibliographystyle{apj}
%\bibliography{reference} 

\begin{table*}
\caption{Properties of Telescopes and Instruments}
\noindent
\begin{tabular}{lll} \hline
Site:            &ARIES Nainital                  & PRL Mount Abu                \\\hline
Telescope:       &1.04-m RC Cassegrain            & 1.20 m Cassegrain            \\
CCD model:       & Wright 2K CCD                  & Andor EMCCD                   \\
Chip size:       & $2048\times2048$ pixels        & $2048\times2048$ pixels       \\
Pixel size:      &$24\times24$ $\mu$m             & $13\times13$ $\mu$m           \\
Scale:           &0.37\arcsec/pixel               & 0.17\arcsec/pixel            \\
Field:           & $13\arcmin\times13\arcmin$     & $3\arcmin\times 3\arcmin$      \\
Gain:            &10 $e^-$/ADU                    & 5 $e^-$/ADU                     \\
Read Out Noise:  &5.3 $e^-$ rms                   & 4.9 $e^-$ rms                    \\
Binning used:    &$2\times2$                      & $2\times2$                      \\
Typical seeing : & 1\arcsec to 2.8\arcsec         & 1\arcsec to 2.6\arcsec           \\\hline
\end{tabular} \\
\noindent
\end{table*}

\begin{table*}
\caption{ Observation Log }
\begin{tabular}{l c c c c c } \hline
Blazar Name & Date of Observation & Telescope  &Filter   & Data Points  & Duration (hrs) \\\hline
3C 66A      &  2008  Oct  22    &A          &R          &68   & 3.86   \\
            &  2008  Oct  26    &A          &R          &72   & 3.50   \\
            &  2008  Dec  23    &B          &R          &90   & 1.45   \\
            &  2008  Dec  24    &B          &R          &113  & 1.85   \\
            &  2008  Dec  27    &B          &R          &417  & 3.46   \\
            &  2008  Dec  28    &B          &R          &203  & 2.23   \\
            &  2009  Jan  03    &B          &R          &320  & 3.54   \\
AO 0235$+$164 &  2008  Oct  20    &A          &R          &98   & 6.30   \\
            &  2008  Oct  23    &A          &R          &45   & 2.60   \\
            &  2008  Dec  26    &B          &R          &248  & 1.23   \\
PKS 0420$-$014&  2008  Oct  23    &A          &R          &18   & 1.94    \\
            &  2008  Dec  26    &B          &R          &29   & 0.73    \\
S5 0716+714 &  2008  Oct  24    &A          &R          &25   & 1.35    \\
            &  2008  Dec  23    &B          &R          &114  & 0.40   \\
            &  2008  Dec  24    &B          &R          &292  & 1.62    \\
            &  2009  Jan  03    &B          &R          &2685 & 3.73    \\
PKS 0735$+$178&  2008  Dec  23    &B          &R          &90   & 0.63   \\
            &  2008  Dec  28    &B          &R          &300  & 3.20   \\
            &  2009  Jan  04    &B          &R          &50   & 1.02   \\
            &  2009  Jan  20    &A          &R          &67   & 3.98   \\
OJ 287      &  2008  Dec  26    &B          &R          &70   & 0.57   \\
            &  2009  Jan  20    &A          &R          &60   & 3.87   \\
            &  2009  Jan  22    &A          &R          &70   & 3.35   \\
3C 273      &  2008  Dec  23    &B          &R          &130  & 0.72    \\
            &  2009  Jan  22    &A          &R          &90   & 3.88    \\
            &  2009  Feb  25    &A          &R          &101  & 3.61    \\   
            &  2009  Mar  24    &A          &R          &36   & 1.38    \\ 
            &  2009  Apr  01    &A          &R          &110  & 5.60    \\
            &  2009  Apr  18    &A          &R          &91   & 3.54    \\
            &  2009  Apr  19    &A          &R          &195  & 7.42    \\
            &  2009  Apr  27    &A          &R          &66   & 3.31    \\
PKS 1510$-$089    &  2009  Apr  17    &A          &R          &77   & 3.82  \\
            &  2009  Apr  19    &A          &R          &22   & 1.09  \\
            &  2009  Apr  27    &A          &R          &64   & 4.47  \\
            &  2009  June 21    &A          &R          &68   & 4.41  \\
BL Lac      &  2008  Sep  04    &A          &R          &83   & 5.10    \\
            &  2008  Oct  26    &A          &R          &73   & 4.25    \\
            &  2009  June 21    &A          &R          &62   & 2.98  \\
3C 454.3    &  2008  Oct  24    &A          &R          &55   & 3.01  \\
            &  2008  Oct  28    &A          &R          &65   & 4.25  \\\hline

\end{tabular}     \\
A : 1.04 m Sampuranand Telescope, ARIES, Nainital, India \\
B : 1.20 m Telescope, PRL, Mount Abu, India 
\end{table*}

\begin{table*}
\caption{ Standard Stars in the  Blazar Fields }
\begin{tabular}{lcccccl} \hline 
Source      & Standard   &  R magnitude    & Refrences$^a$    \\
Name        & star       &  (error)        &              \\\hline
3C 66A      & 1          & 13.36(0.01)    & 5            \\
            & 2          & 14.28(0.04)    & 5            \\
            & 3          & 15.46(0.12)    & 5            \\
            & 4          & 12.70(0.04)    & 6            \\
            & 5          & 13.62(0.05)    & 6            \\
AO 0235+164 & 1          & 12.69(0.02)    & 1           \\
            & 2          & 12.23(0.02)    & 1           \\
            & 3          & 12.48(0.03)    & 1           \\
            & 6          & 13.64(0.04)    & 6           \\
            & 8          & 15.79(0.10)    & 1            \\
            & C1         & 14.23(0.05)    & 6            \\ 
PKS 0420$-$014& 1          & 12.09(0.03)    & 4           \\
            & 2          & 12.80(0.02)    & 5           \\
            & 3          & 12.89(0.01)    & 5           \\
            & 4          & 14.47(0.01)    & 5           \\
            & 5          & 14.37(0.03)    & 4           \\
            & 6          & 14.70(0.03)    & 4           \\
            & 7          & 14.91(0.03)    & 4           \\
            & 8          & 15.46(0.03)    & 4           \\
            & 9          & 15.58(0.04)    & 4           \\
S5 0716+714 & 1          & 10.63(0.01)    & 2            \\
            & 2          & 11.12(0.01)    & 2, 3         \\
            & 3          & 12.06(0.01)    & 2, 3         \\
            & 4          & 12.89(0.01)    & 2            \\
            & 5          & 13.18(0.01)    & 2, 3         \\
            & 6          & 13.26(0.01)    & 2, 3         \\
            & 7          & 13.32(0.01)    & 2            \\
            & 8          & 13.79(0.02)    & 2            \\
PKS 0735+178& A          & 13.14(0.05)    & 1            \\
            & C          & 13.87(0.06)    & 1            \\
            & D          & 15.45(0.06)    & 1            \\
OJ 287      & 2          & 12.46(0.05)    & 1            \\
            & 4          & 13.72(0.06)    & 1             \\
            & 10         & 14.26(0.06)    & 1            \\
            & 11         & 14.67(0.07)    & 1            \\
3C 273      & C          & 11.30(0.04)    & 1            \\
            & D          & 12.31(0.04)    & 1            \\
            & E          & 12.27(0.05)    & 1            \\
            & G          & 13.16(0.05)    & 1            \\
PKS 1510$-$089& 1          & 11.23(0.03)    & 5           \\
            & 2          & 12.95(0.03)    & 5           \\
            & 3          & 13.98(0.09)    & 5           \\
            & 4          & 14.34(0.05)    & 5           \\
            & 5          & 14.35(0.05)    & 4           \\
            & 6          & 14.61(0.02)    & 4           \\
BL Lac      & B           & 11.93(0.05)   & 1            \\
            & C           & 13.69(0.03)   & 1            \\
            & H           & 13.60(0.03)   & 1            \\
            & K           & 14.88(0.05)   & 1            \\
3C 454.3    & A           & 15.32(0.09)   & 6, 9        \\
            & B           & 14.73(0.05)   & 6, 9        \\
            & C           & 13.98(0.02)   & 9, 10       \\
            & D           & 13.22(0.01)   & 9, 10       \\
            & E           & 14.92(0.08)   & 6, 9        \\
            & F           & 14.83(0.03)   & 4, 9         \\
            & G           & 14.83(0.02)   & 4, 9         \\
            & H           & 13.10(0.04)   & 6, 9         \\
            & C1          & 15.27(0.06)   & 6            \\\hline

\end{tabular} \\
$^a$1. \citep{smith1985}; 2. \citep{villata1998}; 3. \citep{ghisellini1997}; 4. \citep{raiteri1998}; 5. \citep{smith1998};
6. \citep{fiorucci1996}; 7. Craine E.R.:Handbook of Quasistellar and BL Lacertae Objects, \citep{angione1971}; 
8. http://www.lsw.uni-heidelberg.de/projects/extragalactic/charts/2251+158.html 
\end{table*}

\begin{table*}
\caption{ IDV Results }
\begin{tabular}{l c c l l l l r r } \hline
Blazar Name & Date of            & C - Test  &\multicolumn{2}{c}{F - Test}    &V            & A ($\%$)  &  \multicolumn{2}{c}{t$_{v}$}     \\        
            & Observation        & value     & F-value     & p-value          &             &           &  SF (hrs)      &DCF (hrs)        \\\hline
3C 66A      &  2008  Oct  22    &3.59        &3.65        & 3.1e$^{-7}$       & V           & 8.1       & $>$3.86       & $>$3.86         \\     
            &  2008  Oct  26    &2.65        &7.03        & 1.9e$^{-14}$      & V           & 5.7       & $>$3.50       & $>$3.50         \\
            &  2008  Dec  23    &0.78        &0.41        & 0.06              & NV          &           &               &         \\
            &  2008  Dec  24    &3.31        &0.61        & 0.02              & NV          &           &               &         \\
            &  2008  Dec  27    &0.88        &0.58        & 0.26              & NV          &           &               &         \\
            &  2008  Dec  28    &0.75        &0.37        & 0.3               & NV          &           &               &         \\
            &  2009  Jan  03    &0.83        &0.63        & 0.35              & NV          &           &               &         \\
AO 0235$+$164 &  2008  Oct  20    &8.32        &76.84       & 2.2e$^{-16}$      & V           & 13.4      & 5.20?          & $>$6.30        \\
            &  2008  Oct  23    &3.23        &10.53       & 1.4e$^{-14}$      & V           & 4.2       & $>$2.60       & $>$2.60         \\
            &  2008  Dec  26    &1.09        &1.43        & 0.04              & NV          &           &               &         \\
PKS 0420$-$014&  2008  Oct  23    &5.44        &31.44       & 6.6e$^{-15}$      & V     & 14.4       &               &         \\
            &  2008  Dec  26    &5.42        &31.24       & 6.1e$^{-15}$      & V           & 12.3      & 0.12,0.18         & 0.18     \\
S5 0716$+$714 &  2008  Oct  24    &0.92        &1.57        & 0.27              & NV          &           &               &         \\
            &  2008  Dec  23    &3.47        &7.77        & 2.2e$^{-16}$      & V           & 9.1       & 0.0014        & 0.0028   \\
            &  2008  Dec  24    &4.51        &3.07        & 2.2e$^{-16}$      & V           & 14.6      & $>$1.62       & $>$1.62        \\
            &  2009  Jan  03    &3.71        &1.26        & 7.7e$^{-7}$       & V           & 31.6      & 3.28          & $>$3.73        \\
PKS 0735$+$178&  2008  Dec  23    &3.14        &9.49        & 2.2e$^{-16}$      & V    & 11.1      &               &         \\
            &  2008  Dec  28    &4.22        &18.36       & 2.2e$^{-16}$      & V           & 19.3      & 2.90          & 2.30     \\
            &  2009  Jan  04    &2.31        &4.64        & 3.1e$^{-7}$       & V           & 9.8       & 0.58          & 0.60        \\
            &  2009  Jan  20    &1.30        &1.99        & 0.055             & NV          &           &               &         \\
OJ 287      &  2008  Dec  26    &1.11        &1.48        & 0.10              & NV          &           &               &         \\
            &  2009  Jan  20    &0.88        &1.20        & 0.47              & NV          &           &               &         \\
            &  2009  Jan  22    &3.21        &11.13       & 2.2e$^{-16}$      & V           & 7.9       &  $>$3.35      & $>$3.35         \\
3C 273      &  2008  Dec  23    &0.91        &0.73        & 0.08              & NV          &           &               &         \\
            &  2009  Jan  22    &0.81        &0.65        & 0.05              & NV          &           &               &         \\
            &  2009  Feb  25    &0.75        &0.49        & 0.06              & NV          &           &               &         \\
            &  2009  Mar  24    &0.76        &0.50        & 0.05              & NV          &           &               &         \\
            &  2009  Apr  01    &2.59        &2.58        & 1.2e$^{-6}$       & V           & 5.2       &   $>$5.2        &$>$5.60         \\
            &  2009  Apr  18    &1.59        &0.25        & 0.11              & NV          &           &               &         \\
            &  2009  Apr  19    &3.74        &2.22        & 4.7e$^{-8}$       & V           & 4.5       & 6.0          &6.0         \\
            &  2009  Apr  27    &0.86        &0.84        & 0.49              & NV          &           &               &         \\
PKS 1510$-$089    &  2009  Apr  17    &2.67        &2.31        & 0.0003            & V           & 9.3       & $>$3.82       & $>$3.82        \\
            &  2009  Apr  19    &2.72        &7.11        & 3.3e$^{-5}$       & V           & 7.6       & 0.60          &0.50         \\
            &  2009  Apr  27    &1.27        &1.54        & 0.09              & NV          &           &               &         \\
            &  2009  June 21    &4.46        &2.65        & 0.0001            & V           & 8.3       & $>$4.41       & $>$4.41        \\
BL Lac      &  2008  Sep  04    &2.73        &5.71        & 1.0e$^{-13}$      & V           & 8.7       & 2.5          & 2.5        \\
            &  2008  Oct  26    &1.05        &1.56        & 0.06              & NV          &           &               &         \\
            &  2009  June 21    &2.87        &7.34        & 4.9e$^{-13}$      & V           &12.9       & $>$2.98       &$>$2.98         \\
3C 454.3    &  2008  Oct  24    &5.14        &22.95       & 2.2e$^{-16}$      & V           & 16.3      &$>$3.01        & $>$3.01        \\
            &  2008  Oct  28    &1.42        &2.09        & 0.01              & NV          &           &               &          \\\hline

\end{tabular} \\
\end{table*}

\begin{figure*}
\epsfig{figure = 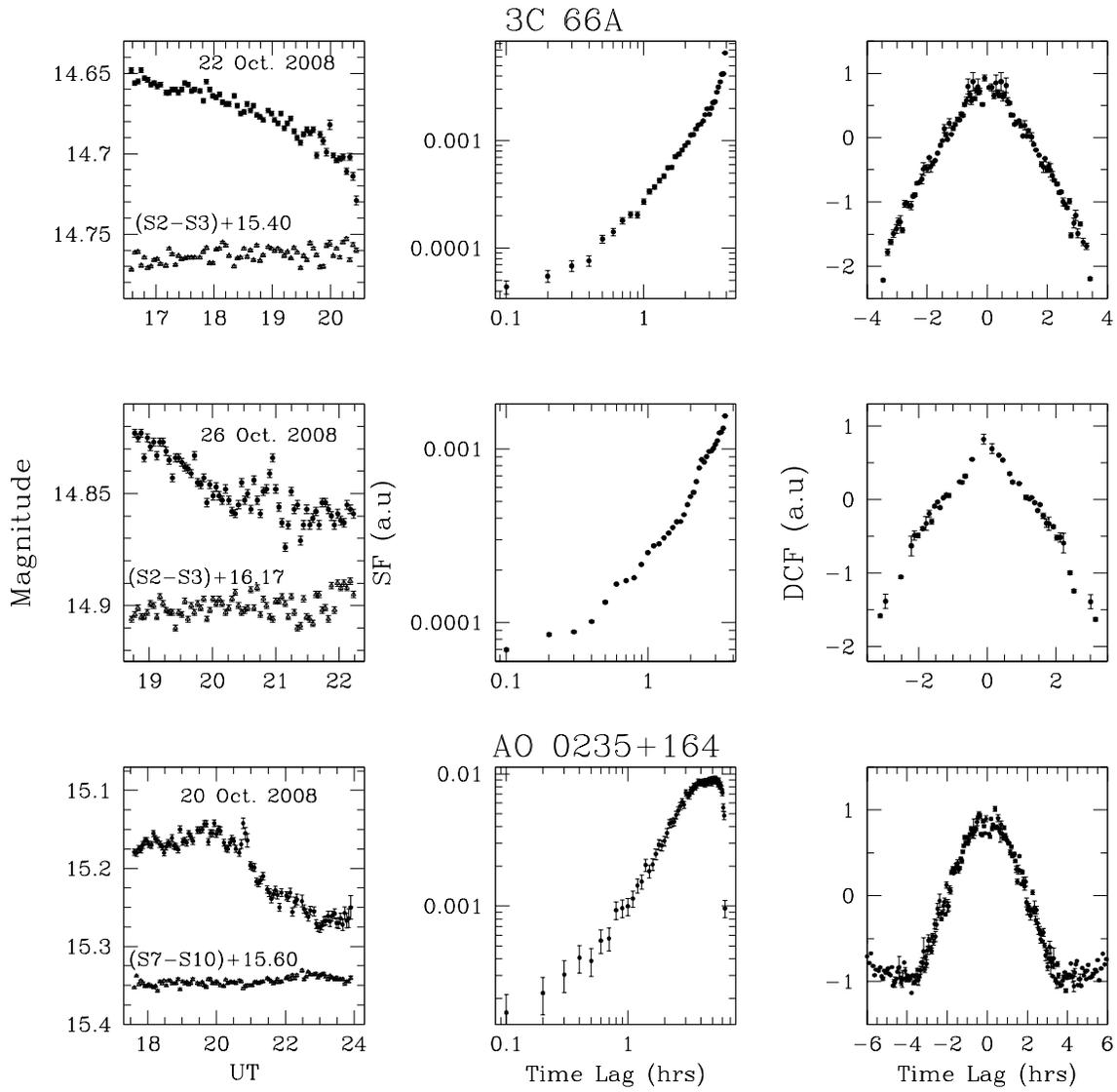,height=16.cm,width=16.cm,angle=0}
\caption{R band optical IDV LCs of the blazars 3C 66A and AO 0235$+$164 and their respective SFs and DCFs.    }
\end{figure*}

\begin{figure*}
\epsfig{figure = 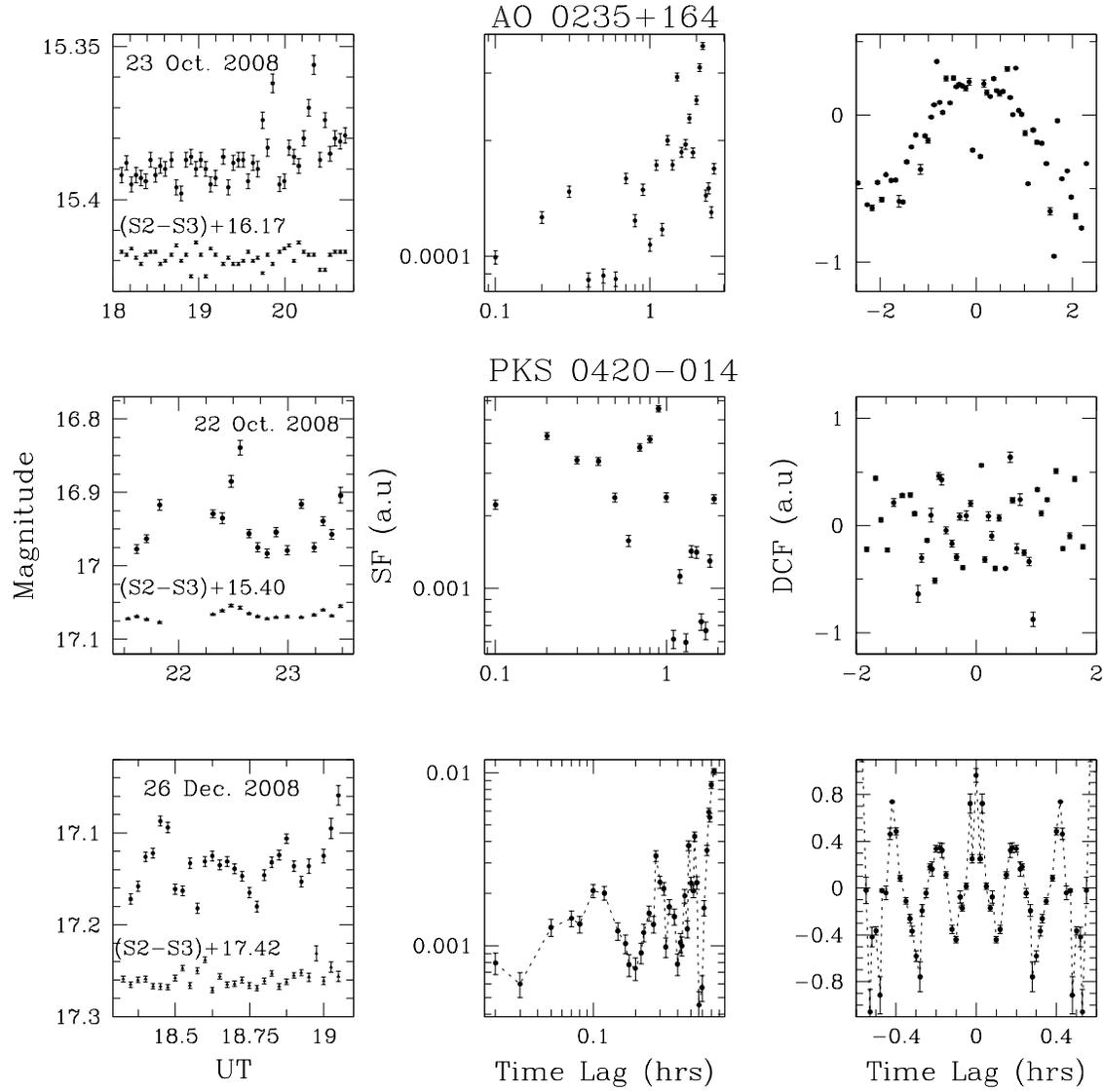,height=16.cm,width=16.cm,angle=0}
\caption{R band optical IDV LCs of the blazars AO 0235$+$164 and PKS 0420$-$014 and their respective SFs and DCFs.         }
\end{figure*}

\begin{figure*}
\epsfig{figure = 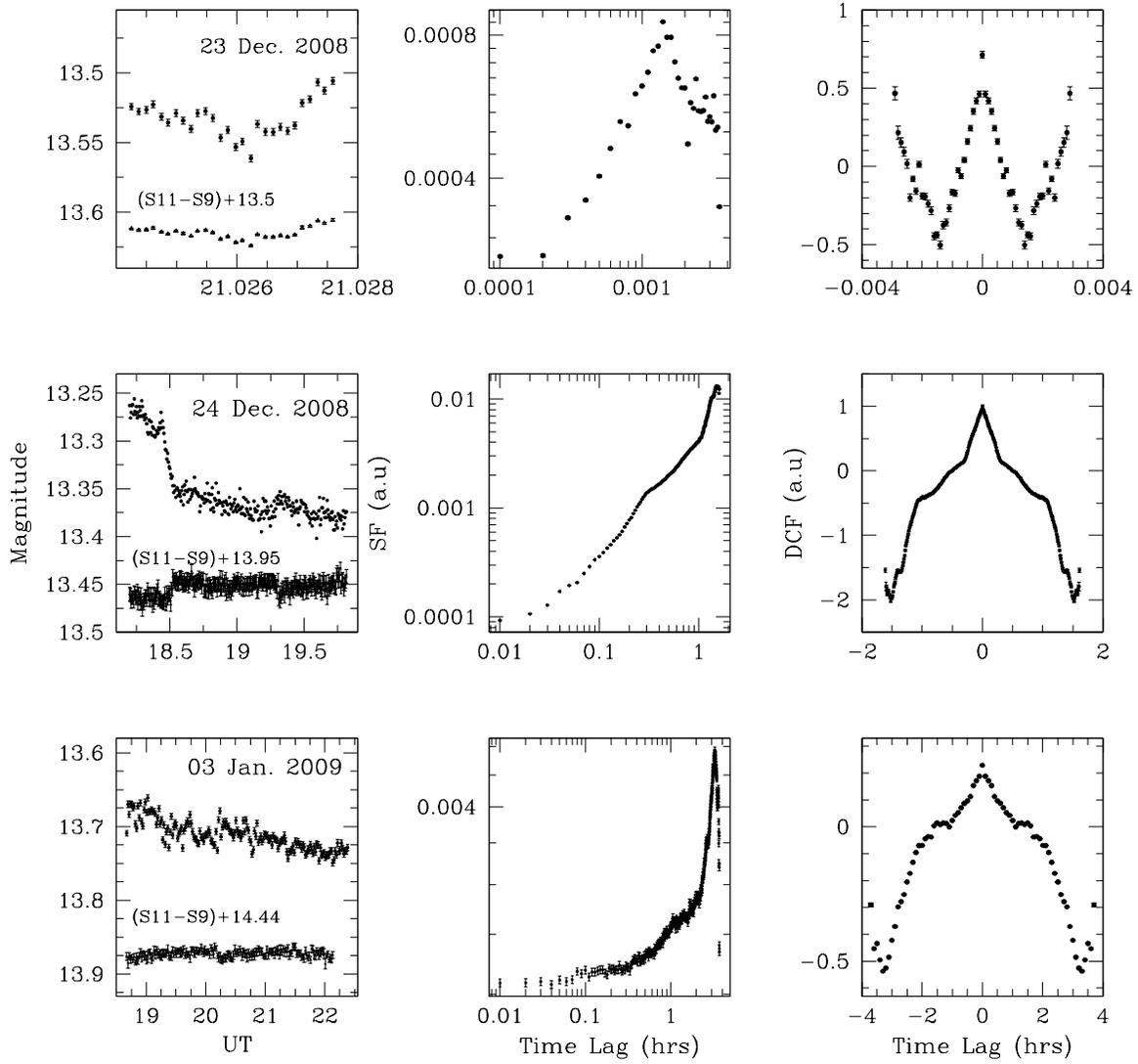,height=16.cm,width=16.cm,angle=0}
\caption{R band optical IDV LCs of the blazar S5 0716$+$714 and their respective SFs and DCFs.}                  
\end{figure*}

\begin{figure*}
\epsfig{figure = 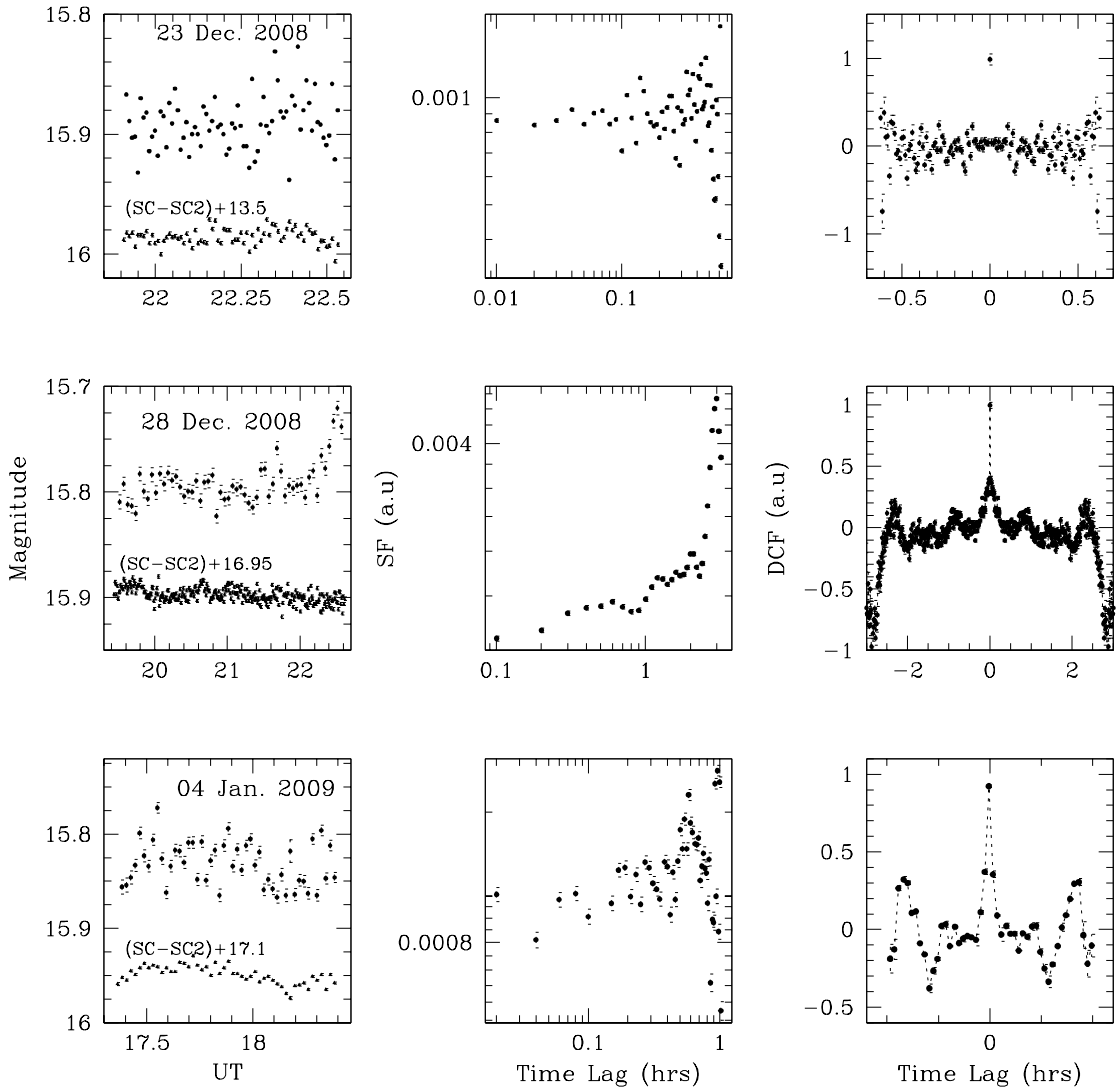,height=16.cm,width=16.cm,angle=0}
\caption{R band optical IDV LCs of the blazar PKS 0735$+$178 and their respective SFs and DCFs.}                 
\end{figure*}

\begin{figure*}
\epsfig{figure = 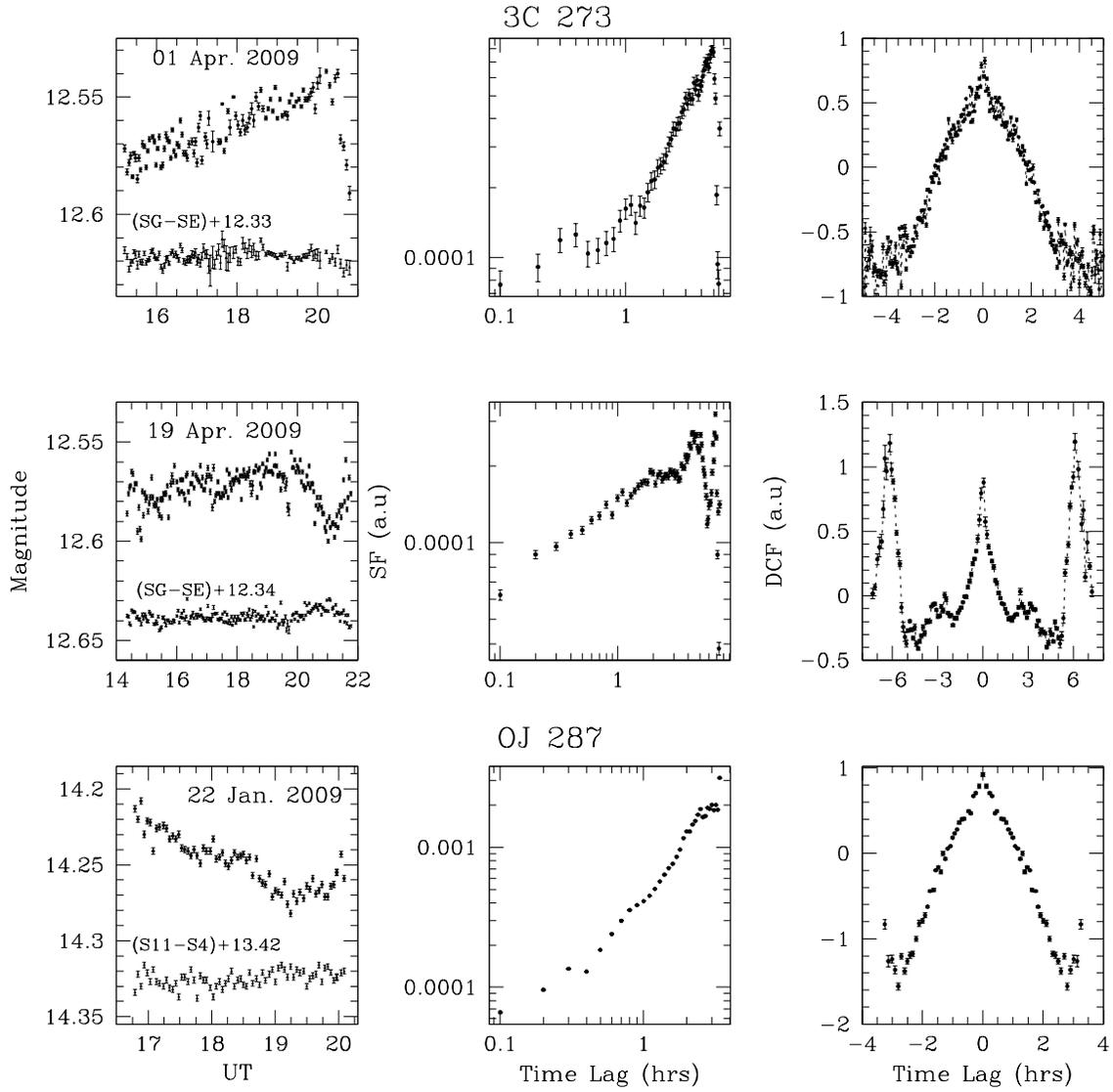,height=16.cm,width=16.cm,angle=0}
\caption{R band optical IDV LCs of the blazars 3C 273 and OJ 287 and their respective SFs and DCFs.          }
\end{figure*}

\begin{figure*}
\epsfig{figure = 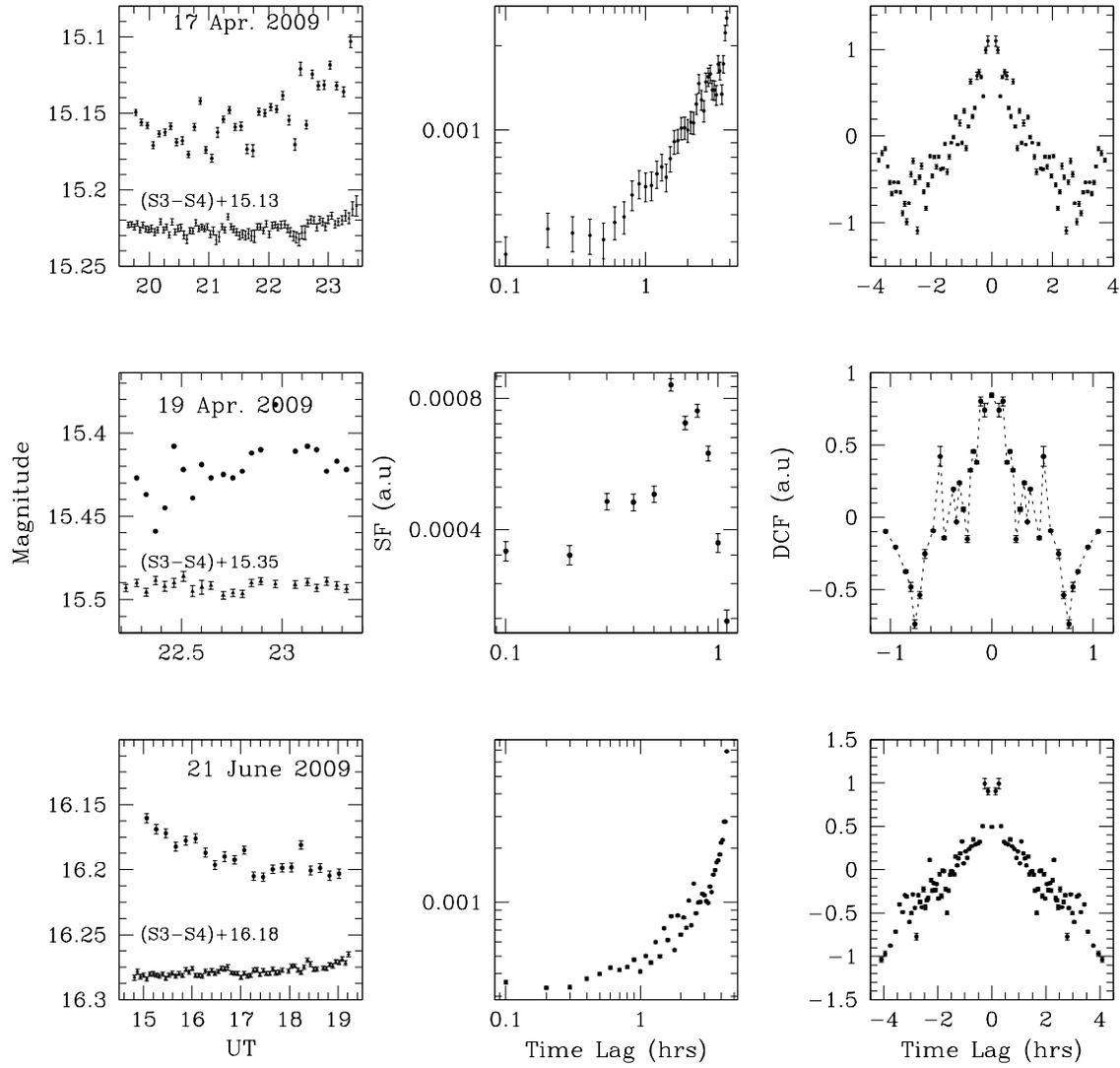,height=16.cm,width=16.cm,angle=0}
\caption{ R band optical IDV LCs of blazar PKS 1510$-$089 and their respective SFs and DCFs.     }
\end{figure*}

\begin{figure*}
\epsfig{figure = 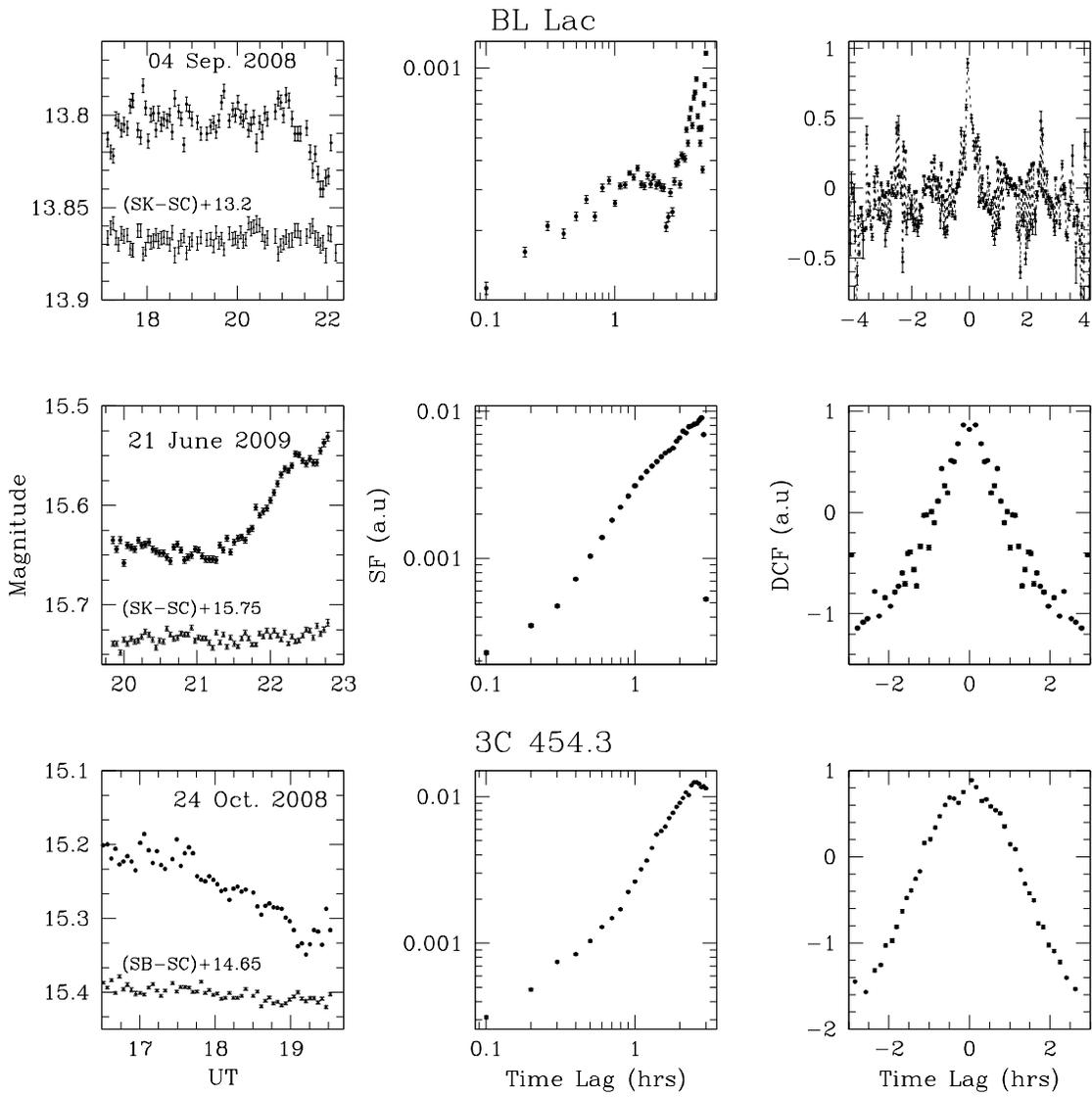,height=16.cm,width=16.cm,angle=0}
\caption{R band optical IDV LC of the blazars BL Lac and 3C 454.3 and their respective SFs and DCFs.}
\end{figure*}

\begin{figure*}
\epsfig{figure = 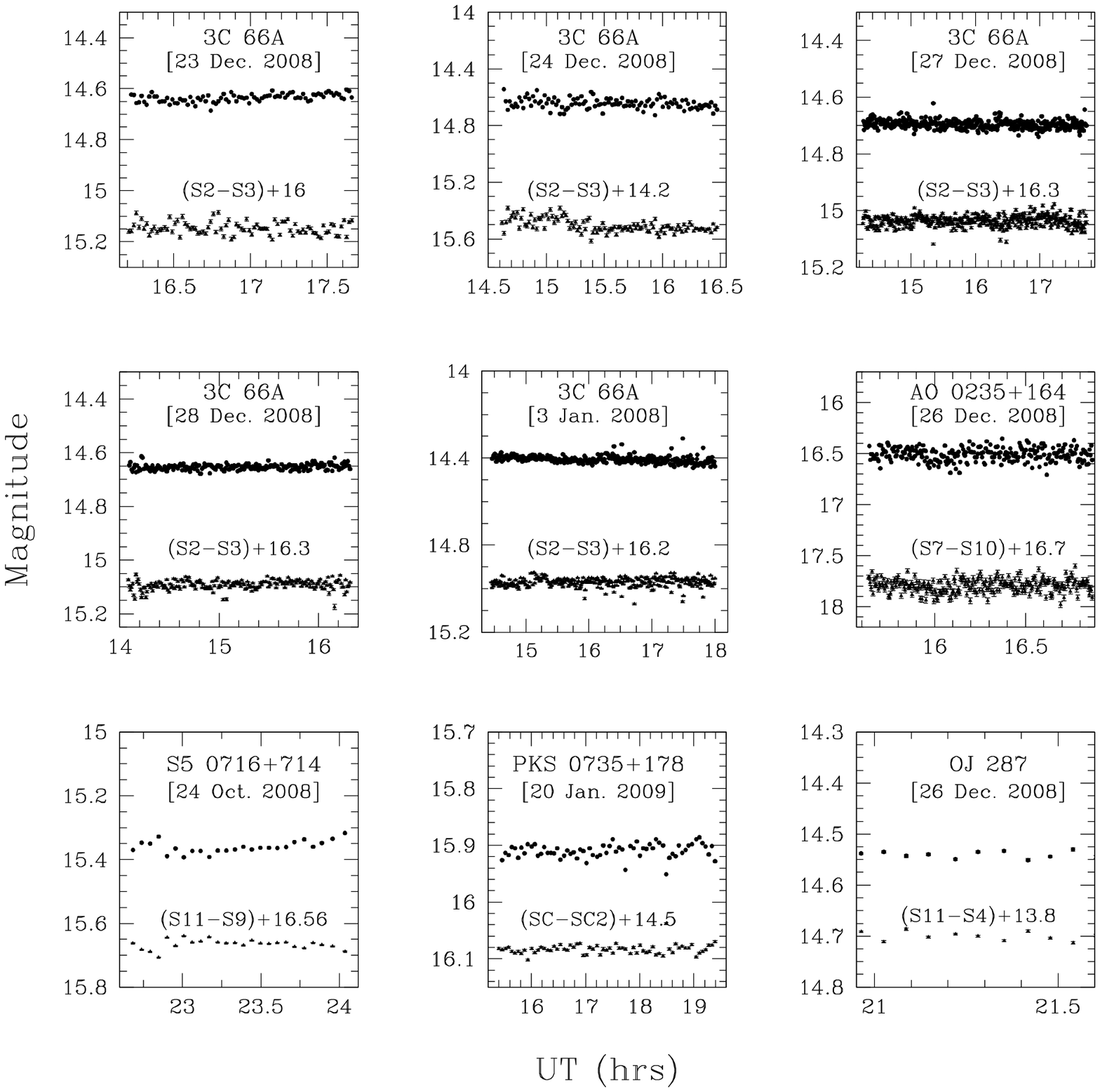,height=16.cm,width=16.cm,angle=0}
\caption{The R filter LCs of several blazars during which no significant variability has been detected, 
plotted along with the differential magnitudes of standard stars.           }
\end{figure*}

\begin{figure*}
\epsfig{figure = 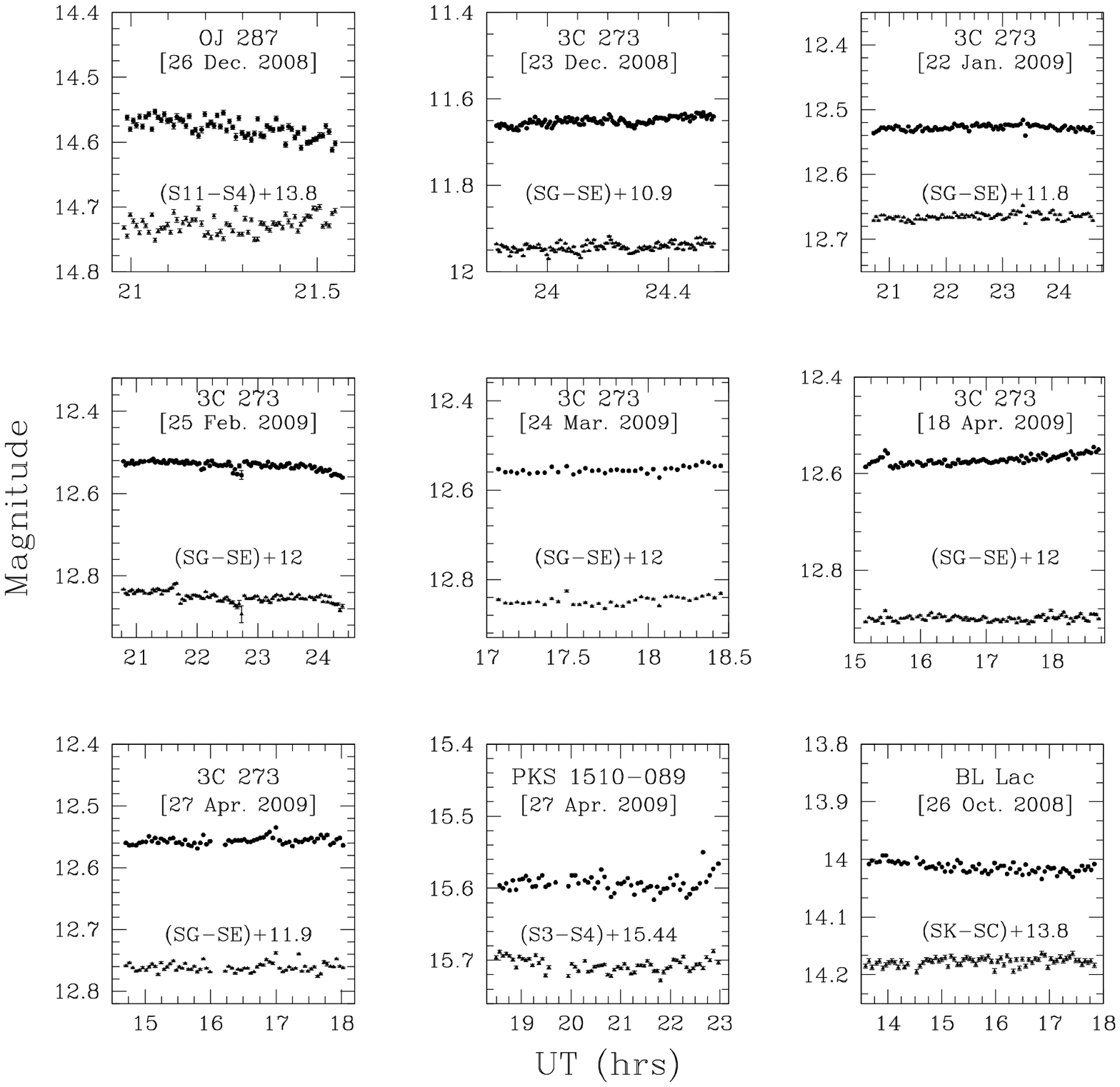,height=16.cm,width=16.cm,angle=0}
\caption{ The R filter LCs of several blazars during which no significant variability has been detected, 
plotted along with the differential magnitudes of standard stars.                 }
\end{figure*}

\label{lastpage}

\end{document}